\documentclass[letterpaper,twocolumn,
prl
,superscriptaddress,email
]{revtex4}

\usepackage{soul}
\usepackage[pdftex]{graphicx}
\usepackage{amssymb}
\usepackage{mathrsfs}
\usepackage{amsmath,amsfonts,latexsym}
\usepackage{array,tabularx,xcolor}
\usepackage{dcolumn} 
\usepackage[normalem]{ulem}
\usepackage{comment}
\usepackage{bm}

\usepackage{doi}
\usepackage{hyperref}

\hypersetup{
    colorlinks=true,
    linkcolor=magenta,
    citecolor=magenta,
    filecolor=magenta,
    urlcolor=red,
}

\begin{document}

\title{Tomography of optical polarization rotation induced by a quantum dot-cavity device}

\author{C. Ant\'on}
\thanks{Equally contributing authors}
\affiliation{Centre de Nanosciences et de Nanotechnologies, CNRS, Univ. Paris-Sud, Universit\'e Paris-Saclay, C2N -- Marcoussis, 91460 Marcoussis, France}

\author{C. A. Kessler}
\thanks{Equally contributing authors}
\affiliation{Centre de Nanosciences et de Nanotechnologies, CNRS, Univ. Paris-Sud, Universit\'e Paris-Saclay, C2N -- Marcoussis, 91460 Marcoussis, France}

\author{P. Hilaire}
\thanks{Equally contributing authors}
\affiliation{Centre de Nanosciences et de Nanotechnologies, CNRS, Univ. Paris-Sud, Universit\'e Paris-Saclay, C2N -- Marcoussis, 91460 Marcoussis, France}
\affiliation{Universit\'e Paris Diderot -- Paris 7, 75205 Paris CEDEX 13, France}

\author{J. Demory}
\affiliation{Centre de Nanosciences et de Nanotechnologies, CNRS, Univ. Paris-Sud, Universit\'e Paris-Saclay, C2N -- Marcoussis, 91460 Marcoussis, France}

\author{C. G\'omez}
\affiliation{Centre de Nanosciences et de Nanotechnologies, CNRS, Univ. Paris-Sud, Universit\'e Paris-Saclay, C2N -- Marcoussis, 91460 Marcoussis, France}

\author{A. Lema\^itre}
\affiliation{Centre de Nanosciences et de Nanotechnologies, CNRS, Univ. Paris-Sud, Universit\'e Paris-Saclay, C2N -- Marcoussis, 91460 Marcoussis, France}

\author{I. Sagnes}
\affiliation{Centre de Nanosciences et de Nanotechnologies, CNRS, Univ. Paris-Sud, Universit\'e Paris-Saclay, C2N -- Marcoussis, 91460 Marcoussis, France}

\author{N. D. Lanzillotti-Kimura}
\affiliation{Centre de Nanosciences et de Nanotechnologies, CNRS, Univ. Paris-Sud, Universit\'e Paris-Saclay, C2N -- Marcoussis, 91460 Marcoussis, France}

\author{O. Krebs}
\affiliation{Centre de Nanosciences et de Nanotechnologies, CNRS, Univ. Paris-Sud, Universit\'e Paris-Saclay, C2N -- Marcoussis, 91460 Marcoussis, France}

\author{N. Somaschi}
\affiliation{Centre de Nanosciences et de Nanotechnologies, CNRS, Univ. Paris-Sud, Universit\'e Paris-Saclay, C2N -- Marcoussis, 91460 Marcoussis, France}

\author{P. Senellart}
\affiliation{Centre de Nanosciences et de Nanotechnologies, CNRS, Univ. Paris-Sud, Universit\'e Paris-Saclay, C2N -- Marcoussis, 91460 Marcoussis, France}

\author{L. Lanco}
\email{loic.lanco@univ-paris-diderot.fr}
\affiliation{Centre de Nanosciences et de Nanotechnologies, CNRS, Univ. Paris-Sud, Universit\'e Paris-Saclay, C2N -- Marcoussis, 91460 Marcoussis, France}
\affiliation{Universit\'e Paris Diderot -- Paris 7, 75205 Paris CEDEX 13, France}

\date{\today}

\begin{abstract}
We introduce a tomography approach to describe the optical response of a cavity quantum electrodynamics device, beyond the semiclassical image of polarization rotation, by analyzing the polarization density matrix of the reflected photons in the Poincar\'e sphere. Applying this approach to an electrically-controlled quantum dot-cavity device, we show that a single resonantly-excited quantum dot induces a large optical polarization rotation by 20$^\circ$ in latitude and longitude in the Poincar\'e sphere, with a polarization purity remaining above 84$\%$. The quantum dot resonance fluorescence is shown to contribute to the polarization rotation via its coherent part, whereas its incoherent part contributes to degrading the polarization purity.
\end{abstract}



\maketitle


In the development of quantum photonic networks, a crucial challenge is to demonstrate highly-efficient interfaces allowing the coherent transfer of quantum information between a stationary qubit and a flying one \cite{Predojevic2015}. In this context, it has been shown that a single natural or artificial atom in a cavity quantum electrodynamics (CQED) system can induce giant phase shifts on incoming photons \cite{Turchette1995,Hofmann2003,Fushman2008,Sames2014}: this allowed the recent implementation of atom-photon gates \cite{Tiecke2014,Reiserer2014,Sun2016} and, subsequently, photon-photon gates \cite{Shomroni2014,Hacker2016}. Most of these achievements are based on polarization-encoding protocols \cite{Duan2004}, in which the optical phase shift is used to rotate the polarization state of the outgoing photons. Polarization rotation is also at the heart of numerous quantum computation proposals based on spin-photon logic gates, where a stationary spin qubit is used as a quantum memory with a long coherence time \cite{Leuenberger2006,Hu2008,Hu2008a,Bonato2010,Smirnov2015}. 

By essence, these concepts of optical phase shift and polarization rotation rely on a semiclassical image, considering that the optical response of the atom-based device is entirely coherent. Yet, the output photonic field includes a contribution from the resonance fluorescence emitted by the natural or artificial atom, which can be partially incoherent with respect to the incoming laser \cite{BookCohenTannoudji}. In quantum optics this is described by an output field operator $\hat{b}$ which cannot be reduced to its expectation value $\langle \hat{b} \rangle$. The coherent contribution is governed by the average field $\langle\hat{b} \rangle$, which keeps a defined phase with respect to the incoming field; the incoherent component, however, arises from quantum fluctuations around the average field, and has no specific phase \cite{BookCohenTannoudji}. A direct consequence of this is that the optical response of a device cannot, in general, be solely described by a reflection coefficient with a well-defined amplitude and phase. Similarly, a general output polarization is not necessarily a pure polarization state. 

The distinction between coherent and incoherent responses is especially important in solid-state quantum technologies, where artificial atoms, such as semiconductor quantum dots (QDs), interact with a fluctuating environment. As was experimentally demonstrated using spectrally-resolved \cite{Matthiesen2012, Bennett2016} or interferometry-based \cite{Nguyen2011, Fischer2016,Bennett2016} measurements, the fluorescence emitted by a resonantly-excited QD has its coherent fraction at best equal to $\frac{T_2}{2T_1}$, with $T_1$ the lifetime of the photon wavepacket and $T_2$ its coherence time \cite{BookCohenTannoudji}. QD-based CQED structures have also been used to demonstrate QD-induced optical phase shifts up to a few tens of degrees \cite{Fushman2008,Young2011,Bakker2015a}, and a spin-dependent polarization rotation of up to $\pm 6^\circ$ \cite{Arnold2015}. None of the reported works, however, have gone beyond the semiclassical image to describe the global optical response of a CQED device, where the response of the cavity itself is superposed to the extracted resonance fluorescence.

Here we introduce a polarization tomography approach to investigate the polarization rotation of coherent light interacting with an electrically-controlled QD-cavity device \cite{Somaschi2016,Giesz2016,DeSantis2016}. We analyze the polarization density matrix in the Poincar\'e sphere \cite{BookBlumDensityMatrix}, which allows distinguishing between a general mixed polarization and a pure polarization state. We show that the superposition of emitted single photons (H-polarized) with reflected photons (V-polarized) leads to a large rotation of the output polarization, by 20$^\circ$ both in latitude and longitude in the Poincar\'e sphere \cite{NotePolarizationEllipse}, with a polarization purity above 84 $\%$. We implement a CQED model with which we analyze the complete information set provided by the polarization tomography. We demonstrate that the coherent part of the QD emission contributes to the polarization rotation, while its incoherent part contributes to degrading the polarization purity. 

\begin{figure}
\setlength{\abovecaptionskip}{-5pt}
\setlength{\belowcaptionskip}{-2pt}
\begin{center}
\includegraphics[width=1\linewidth,angle=0]{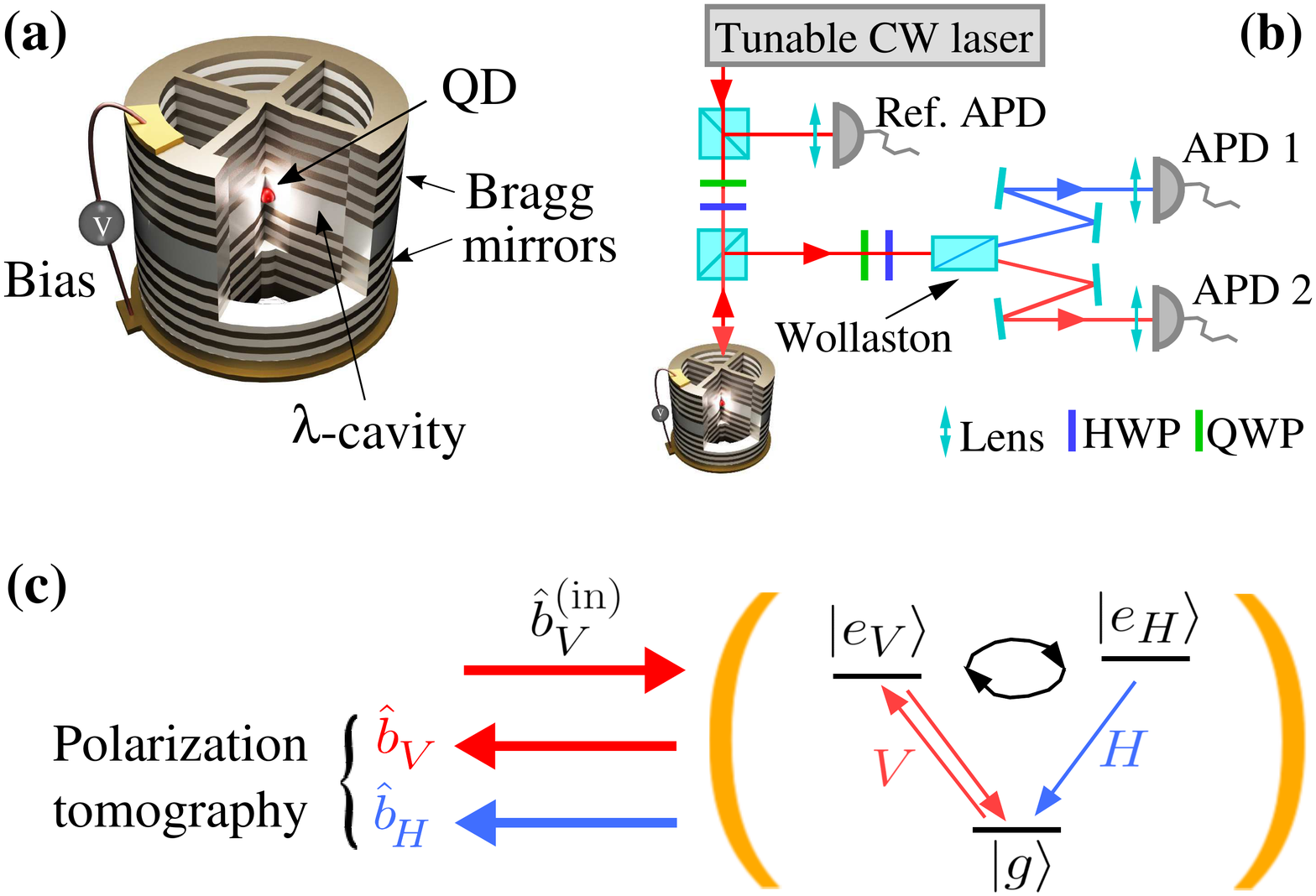}
\end{center}
\caption{\textbf{(a)}~ Schematic representation of the electrically-controlled QD-cavity device. \textbf{(b)}~Sketch of the experimental setup. H(Q)WP: half-(quarter-)wave plate.  Ref. APD denotes the reference photodiode measuring the input field, while APD1 and APD2 are used to analyze the output field. \textbf{(c)} Sketch of the input/output fields and of the embedded three-level system, in the basis of the cavity eigenaxes $V/H$.}
\label{fig:fig1}
\end{figure}

We study a deterministically-coupled, electrically controlled QD-cavity system. The sample, grown by molecular beam epitaxy, consists in a $\lambda$-GaAs cavity embedding InGaAs QDs, surrounded by two distributed Bragg reflectors (GaAs and Al$_{0.9}$Ga$_{0.1}$As, with 30 and 20 pairs for the bottom and top mirrors). In-situ lithography was used to deterministically fabricate a micropillar cavity around a single InGaAs QD \cite{Dousse2008}. The 3~$\mu m$ diameter micropillar is connected by four ridges to a circular frame where metallic ohmic contacts are defined (Fig.~\ref{fig:fig1}(a)). This cavity presents a small anisotropy leading to linearly polarized modes considered as $H$ (horizontal) and $V$ (vertical) polarizations, with a $70$ $\mu$eV splitting. The metallic contacts allow the electrical tuning of the QD transition with respect to these cavity modes \cite{Nowak2014}.

The experimental setup is sketched in Fig.~\ref{fig:fig1}(b): the device is kept inside a helium exchange gas cryostat at $\sim10$\,K, and a tunable continuous-wave laser with 1\,MHz linewidth excites the cavity from the top mirror. The reflected beam can be separated in two orthogonally-polarized components in various polarization bases, using calibrated waveplates and a Wollaston polarizing prism. The input and output field intensities can then be measured with avalanche photodiodes (APDs), as displayed in Fig.~\ref{fig:fig1}(b), or analyzed via spectral or autocorrelation measurements (not shown).

As sketched in Fig.~\ref{fig:fig1}(c), the experiment we perform consists in exciting the device with a $V$-polarized input field (denoted $\hat{b}^{(in)}_V$), corresponding to one of the cavity eigenaxes. In the case of a fully-detuned QD this would imply a $V$-polarized reflected beam; yet, the interaction with a QD optical transition leads to a more complex output. Indeed, the neutral QD can be described as a three-level system with a ground state $|g\rangle$ and two excitonic states $|e_V\rangle$ and $|e_H\rangle$, that can respectively be excited with $V$- and $H$-polarized light. However, $|e_H\rangle$ and $|e_V\rangle$ are not the system eigenstates, as the QD displays an anisotropy along two axes $X$ and $Y$, differing from $H$ and $V$ by an angle $\theta$ \cite{suppmat}. As a consequence, an initially excited state $|e_V\rangle$ coherently oscillates between $|e_H\rangle$ and $|e_V\rangle$ (Fig.~\ref{fig:fig1}(c)). By exciting the system with $V$ polarized light one thus populates the $|e_V\rangle$ state, which rotates into $|e_H\rangle$: this leads to resonance fluorescence emitted both in $V$ and $H$ polarizations. As sketched in Fig.~\ref{fig:fig1}(c), polarization tomography provides a global analysis of the $V$- and $H$-polarized output fields, denoted $\hat{b}_V$ and $\hat{b}_H$. In particular, if both the $H$- and $V$ polarized output fields are coherent with the incoming laser, they superpose with a well-defined phase, resulting in a pure polarization state.

\begin{figure}
\setlength{\abovecaptionskip}{-5pt}
\setlength{\belowcaptionskip}{-2pt}
\begin{center}
\includegraphics[width=1\linewidth,angle=0]{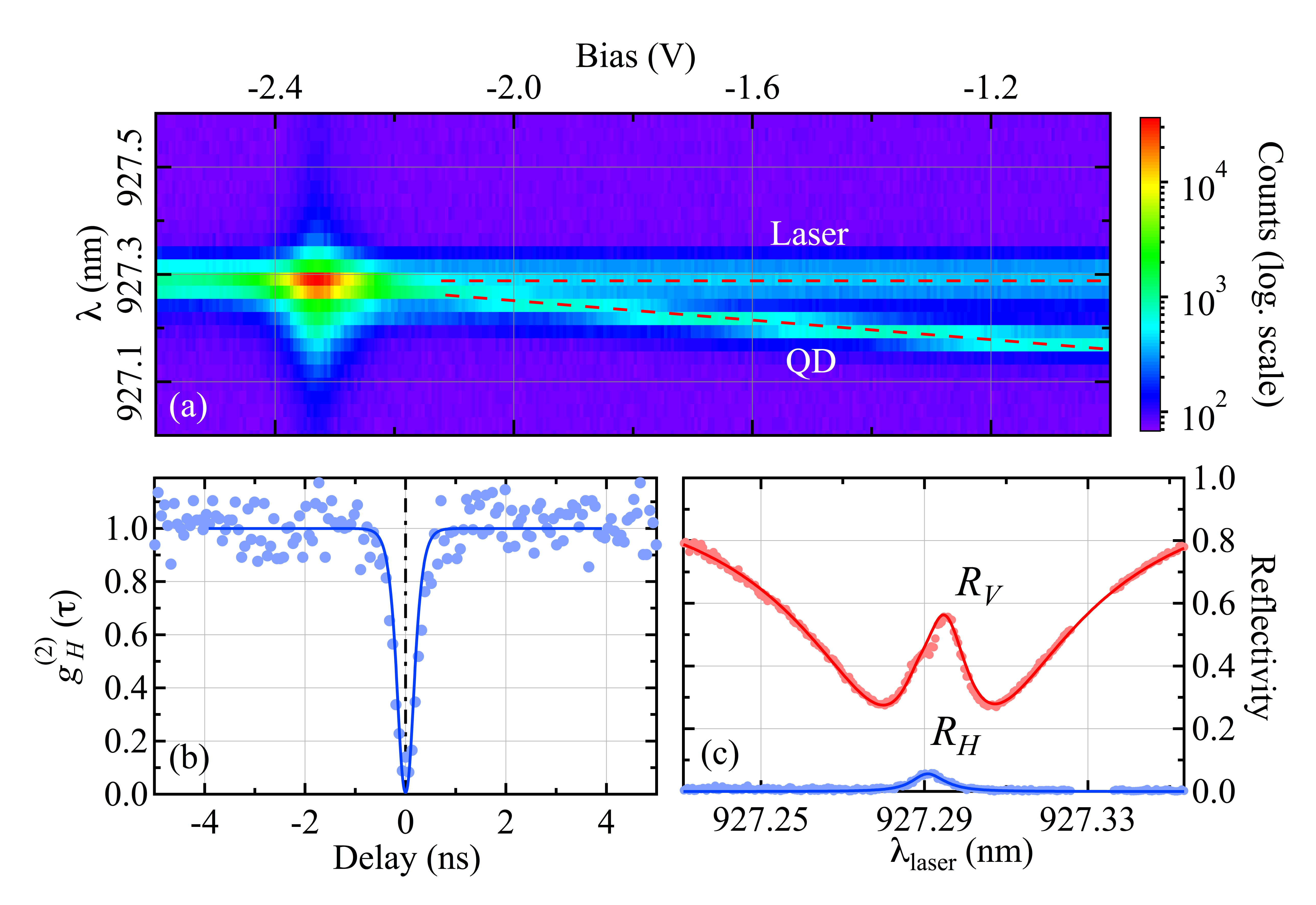}
\end{center}
\caption{\textbf{(a)}~Map of the QD $H$-polarized resonance fluorescence as function of the bias voltage and emission energy, measured for $\lambda_{laser}=927.29$~nm. \textbf{(b)}~Autocorrelation measurements of the $H$-polarized signal, measured for $\lambda_{laser}=\lambda_{QD}=927.29$~nm, and corresponding numerical prediction taking into account the finite detector response time \textbf{(c)}~Normalized intensities in $H$ and $V$ polarizations, measured for $\lambda_{QD}=927.29$~nm and a varying $\lambda_{laser}$, together with numerical predictions.}
\label{fig:fig2}
\end{figure}

We first analyze the device in the cavity eigenbasis $H/V$. Setting the polarization analyzer to separate the $H$ and $V$ polarizations, we measure the spectrum of the $H$-polarized output field with a spectrometer coupled to a CCD camera, filtering out the $V$-polarized light. Fig. \ref{fig:fig2}(a) displays the $H$-polarized signal as a function of the bias voltage applied to the device and of the emission wavelength, for a fixed laser wavelength $\lambda_{laser}=927.29$~nm (in resonance with the bias-independent $V$-polarized cavity mode) and a fixed incoming power $P_{in}=200$~pW (in the low-power regime where no saturation of the QD excitonic transition is observed). When the bias is tuned to $-2.33\,$V, which tunes the QD transition wavelength $\lambda_{QD}$ in resonance with $\lambda_{laser}$, the emitted intensity is strongly increased as expected for a resonance fluorescence process. The logarithmic scale used in Fig. \ref{fig:fig2}(a) also allows observing two residual features at biases above $-2$V: one corresponding to residual light at the laser wavelength, and one corresponding to Raman-assisted QD emission. 

The $H$-polarized signal arises from cross-polarized resonance fluorescence, i.e. from single photons emitted by the resonantly-driven QD. This is evidenced by the measurement of the second-order autocorrelation of the $H$-polarized field, using a standard Hanbury-Brown-Twiss experiment with two single-photon avalanche diodes \cite{Michler2000}, with $\lambda_{laser}=\lambda_{QD}$ both fixed. Fig.~\ref{fig:fig2}(b) displays the corresponding histogram of $g_H^{\left(2\right)}\left(\tau\right)$, with $\tau$ the delay between photon detection events. The raw value of $g_H^{\left(2\right)}\left(0\right)$ decreases down to $7 \pm 5\%$, which is compatible with a single-photon emission by the $|e_H\rangle$ state if one takes into account the finite time response of the single-photon diodes. As is also displayed in Fig.~\ref{fig:fig2}(b), a good agreement is found between the experimental data and our numerical prediction, obtained with CQED parameters that will be discussed later on.



We then measure the intensities of the $V$- and $H$-polarized components of the output field, denoted $I_V$ and $I_H$. In the following we work at a fixed bias of $-2.33$~V, fixing $\lambda_{QD}=927.29~nm$ to be in resonance with the $V$-polarized cavity mode, while scanning the laser wavelength. Fig.~\ref{fig:fig2}(c) displays the evolution of the normalized intensities $I_V/I_{in}$ and $I_H/I_{in}$, with $I_{in}$ the input field intensity, as a function of $\lambda_{laser}$. A peak in $I_H/I_{in}$ is observed at $\lambda_{laser}=\lambda_{QD}$, indicating that up to 7$\%$ of the incident light has been reconverted by the quantum dot into $H$-polarized resonance fluorescence. The $V$-polarized output intensity also presents a peak at $\lambda_{QD}$, though superposed to the broader reflectivity dip of the $V$-polarized cavity mode. This behavior is understood by considering that the $V$-polarized light arises from the superposition of two fields: light directly reflected upon the top-mirror surface, and light extracted from the cavity via the top-mirror \cite{Loo2012}. The high QD-induced peak is thus related to the $V$-polarized resonance fluorescence, emitted by the  $|e_V\rangle$ exciton, which is large enough to strongly change the amount of intracavity light, and thus the resulting output field.

To perform a complete polarization tomography we now analyse our device in the other two bases. By adjusting the waveplates of the polarization analyzer introduced in Fig.~\ref{fig:fig1}(b), we measure the intensities $I_D$ and $I_A$ in the diagonal/anti-diagonal polarization basis, and $I_R$ and $I_L$ in the right-handed/left-handed circular polarization basis. For a given set of intensities $I_{\parallel/\perp}$, we define the corresponding Stokes component as $s_{\parallel\perp}=\left(I_\parallel-I_\perp\right)/\left(I_\parallel+I_\perp\right)$. This allows measuring the density matrix of the polarization state, and representing it in the Poincar\'e sphere as a vector whose coordinates are $s_{HV}$, $s_{DA}$ and $s_{RL}$, ranging between $-1$ and $+1$. The purity of the polarization density matrix is given by the norm of the Poincar\'e vector, $\sqrt{s_{HV}^2+s_{DA}^2+s_{RL}^2}$. This norm is equal to 1 only for a pure polarization state, as would be given by a coherent superposition of $H$- and $V$-polarized electromagnetic fields. 

\begin{figure}
\setlength{\abovecaptionskip}{-5pt}
\setlength{\belowcaptionskip}{-2pt}
\begin{center}
\includegraphics[width=1\linewidth,angle=0]{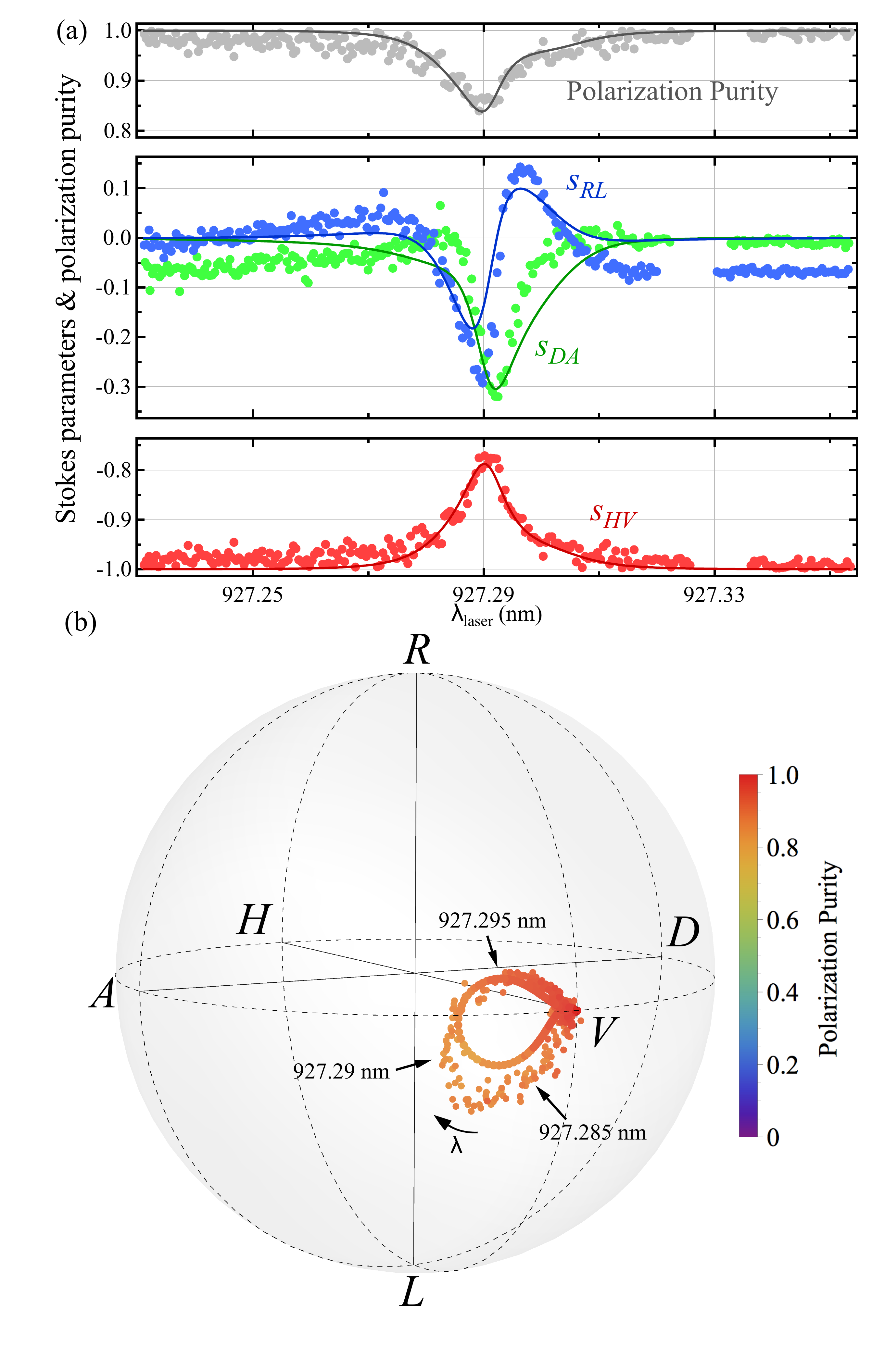}
\end{center}
\caption{(a)~Stokes parameters and polarization purity as a function of $\lambda_{\textrm{laser}}$, using the experimental conditions of Fig.\ref{fig:fig2}(c). Markers: experimental data. Solid lines: numerical fit (see text). (b)~Representation of the polarization state in the Poincar\'e sphere for both experimental data (circles) and numerical simulations (continuous line). Each point is colored according to its corresponding polarization purity.}
\label{fig:fig3}
\end{figure}


In Fig.~\ref{fig:fig3}(a) the three Stokes components $s_{HV}$, $s_{DA}$ and $s_{RL}$, as well as the polarization purity $\sqrt{s_{HV}^2+s_{DA}^2+s_{RL}^2}$, are displayed as a function of $\lambda_{laser}$. The same set of data is also illustrated in Fig.~\ref{fig:fig3}(b), where the Stokes components are used as three dimensional Cartesian coordinates on the Poincar\'e sphere, for the various values of $\lambda_{laser}$. Far from the QD resonance we obtain $s_{HV}\approx-1$ together with $s_{DA}\approx 0$ and $s_{LR}\approx 0$. This corresponds to a $V$-polarized reflected field, as also illustrated by the accumulation of experimental points at the $V$ polarization in the Poincare sphere. At resonance a maximal value of $s_{HV}=-0.77$ is obtained, and at the same time the Stokes component $s_{DA}$ and $s_{RL}$ become non-zero in the region of the QD resonance. As seen in the Poincar\'e sphere this corresponds to a rotation between 0 and more than 20 degrees, both in longitude and latitude, when $\lambda_{laser}$ is tuned around $\lambda_{QD}=927.29$ nm. 

These results, translated in terms of angles in the polarization ellipse, correspond to a rotation of the orientation and ellipticity angles of more than 10 degrees. They do not mean, however, that the output polarization is a pure polarization state: indeed a degradation of the polarization purity is observed around the QD resonance, though the purity remains above 84$\%$ (Fig.~\ref{fig:fig3}(a), top panel). This is also illustrated in the Poincar\'e sphere of Fig.~\ref{fig:fig3}(b), where each experimental point is colored following a scale indicating the corresponding polarization purity, i.e. the distance between the data point and the center of the sphere.

We now analyze our experimental results based on the model sketched in Fig.~\ref{fig:fig1}(c), using the input-output formalism \cite{Walls2008,Auffeves-Garnier2007} with a single input operator $\hat{b}^{(in)}_V$ and two output operators $\hat{b}_V$ and $\hat{b}_H$ \cite{suppmat}. This allows computing the intensities $I_{\parallel}$ and $I_{\perp}$ in any polarization basis, which gives access to the three Stokes components and to the expected polarization purity. Fitting simultaneously the experimental curves in Figures \ref{fig:fig2}(c) and \ref{fig:fig3}(a) allows determining all the parameters at play, more precisely than what was previously possible with the analysis of optical nonlinearities in QD-cavity devices \cite{Loo2012,DeSantis2016}. We first extract a QD-cavity light-matter coupling $g=18\pm3$ $\mu$eV, a total cavity damping rate $\kappa=106\pm 4$ $\mu$eV for both $H$ and $V$ modes, and a total QD decoherence rate $\gamma=4\pm 0.5$ $\mu$eV (using $\hbar=1$ units). The QD anisotropy is also found to be characterized by a fine structure splitting $\Delta_{\textrm{FSS}}=9\pm 2$ $\mu$eV between the excitonic eigenstates $|e_X\rangle$ and $|e_Y\rangle$, with the $X/Y$ axes rotated by an angle $\theta=17\pm 5^\circ$ with respect to the $H$/$V$ cavity axes. The top-mirror output-coupling $\eta_{top}$, which is the ratio between the top-mirror damping rate $\kappa_{top}$ and the total damping rate $\kappa$, is best fitted at $\eta_{top}=55\pm 5\%$. 


If the QD resonance fluorescence field were entirely coherent with respect to the incoming laser, the output light would be completely described by only two expectation values $\langle\hat{b}_{H}\rangle$ and $\langle\hat{b}_{V}\rangle$, each having a well defined amplitude and phase: this would lead to a coherent, classical superposition of $H$- and $V$- polarized fields, i.e. a pure polarization state. Yet, partial incoherence has to be considered to interpret the reduced polarization purity: in the numerical fits of Fig.~\ref{fig:fig3}(a), this partial incoherence arises from the residual pure dephasing experienced by the exciton, whose rate is estimated at $\gamma^*=3.7\pm0.5$ $\mu$eV. We attribute this dephasing to fluctuations of the QD transition energy, which could be induced by a residual current at the bias of $-2.33$V, and/or by fluctuations of the helium flow in our exchange gas cryostat. Taking into account that the total QD decoherence rate $\gamma$ is given by $\gamma=\frac{\gamma_{sp}}{2}+\gamma^*$, with $\gamma_{sp}$ the rate of spontaneous emission in the external (leaky) optical modes, we determine that $\gamma_{sp}=0.6$ $\mu$eV, consistent with a 1 ns leak emission time. We also note that the same set of parameters has been used to provide the numerical predictions in Fig. 2(b), without any additional fitting, in agreement with the measured autocorrelation function.


\begin{figure}
\setlength{\abovecaptionskip}{-5pt}
\setlength{\belowcaptionskip}{-2pt}
\begin{center}
\includegraphics[width=1\linewidth,angle=0]{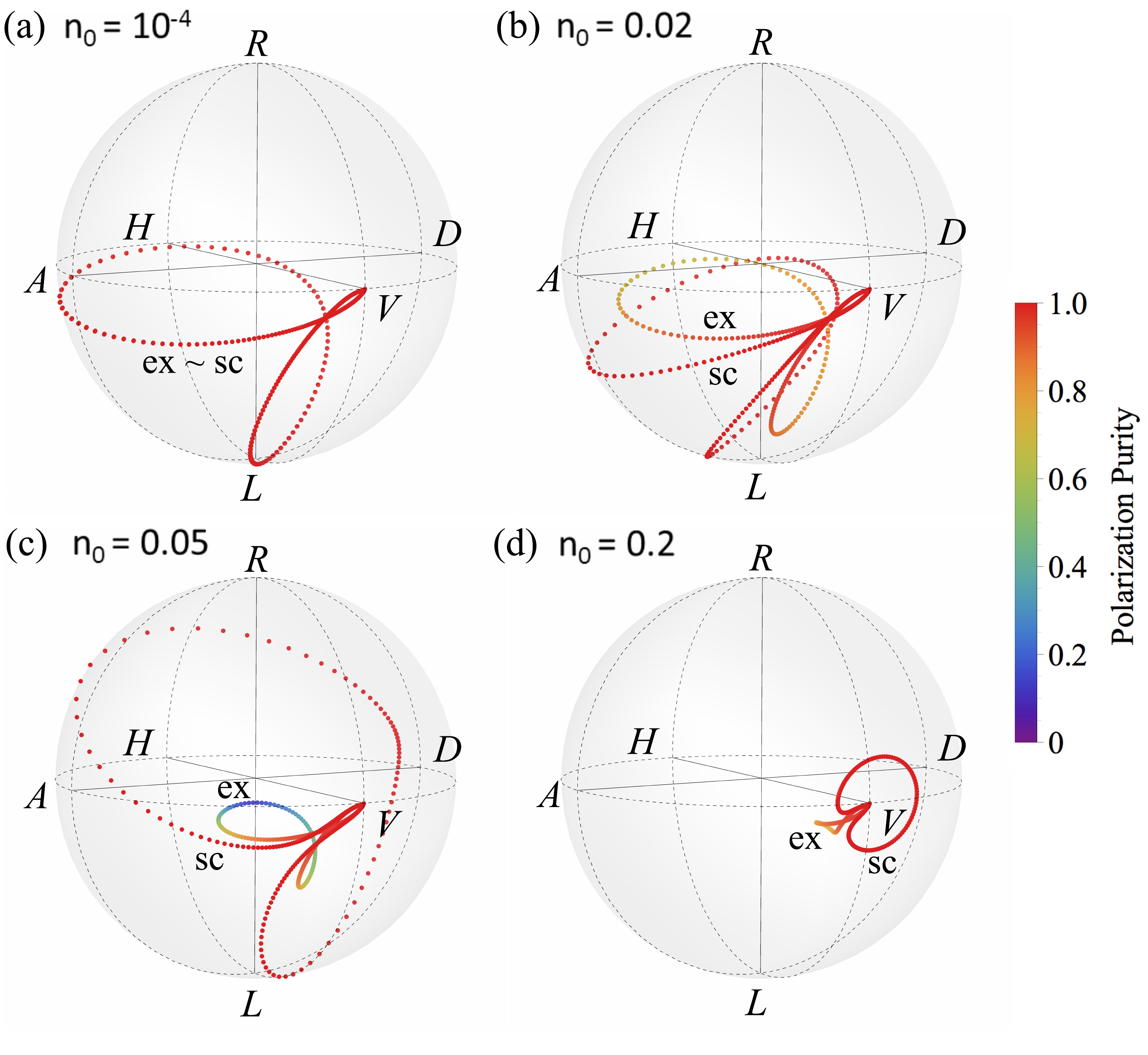}
\end{center}
\caption{Theoretical calculation of the Stokes parameters represented in the Poincar\'e sphere for various values of $\lambda_{\textrm{laser}}$. Same parameters as those used in Fig. \ref{fig:fig3} except $\gamma^*=0$, $\theta=45^\circ$ and $\eta_{top}=1$. Each panel corresponds to a different intracavity photon number $n_0$, where $n_0$ is the intracavity photon number that is obtained at resonance in the empty-cavity regime, deduced from $n_0=\frac{4 P_{in}}{\kappa \hbar \omega_c}$, and contains two curves corresponding to the exact calculation (ex), and to the semiclassical approximation (sc). Each point is colored according to its corresponding polarization purity.}
\label{fig:fig4}
\end{figure}

Pure dephasing, however, is not the only source of fluctuations leading to a partially incoherent output: quantum fluctuations around the expectation values $\langle\hat{b}_V\rangle$ and $\langle\hat{b}_H\rangle$ are also obtained in the intermediate-power regime, i.e. when the excitation results in a population of the excitonic transition \cite{Bakker2015}. We theoretically study, in Fig.~\ref{fig:fig4}, the polarization tomography of an optimized device, with a pure dephasing rate $\gamma^*=0$, a top-mirror output-coupling $\eta_{top}=1$ and a QD-cavity eigenaxis angle $\theta=45^\circ$ (the other parameters being unchanged). The calculation is performed for various incoming powers $P_{in}=$1, 100, 500 pW and 2 nW, corresponding to different values of the intracavity photon number $n_0$ indicated in Figs~\ref{fig:fig4}(a-d). In each panel two numerical calculations are compared: the exact one considering that the intensities $I_{\parallel}$ and $I_{\perp}$, in an arbitrary basis, are given by $\langle \hat{b}_{\parallel}^{\dagger}\hat{b}_{\parallel}\rangle$ and $\langle \hat{b}_{\perp}^{\dagger}\hat{b}_{\perp}\rangle$, and the semiclassical one neglecting the incoherent part of the output fields, i.e. considering that $I_{\parallel}=\langle \hat{b}_{\parallel}^{\dagger}\rangle\,\langle\hat{b}_{\parallel}\rangle$ and $I_{\perp}=\langle \hat{b}_{\perp}^{\dagger}\rangle\,\langle\hat{b}_{\perp}\rangle$. At low incoming power, in Fig.~\ref{fig:fig4}(a), the semiclassical approximation gives the same result as the exact calculation. Maximal polarization rotations are obtained and the QD transition even allows converting the $V$-polarized incoming light into a fully $H$-polarized output, as required for a photon-photon gate \cite{Shomroni2014,Hacker2016}. In this situation the output polarization vectors stay at the surface of the sphere, as expected for fully coherent fields. At higher incoming powers, however, the incoherent resonance fluorescence emitted by the quantum dot modifies this situation. As displayed in Figs~\ref{fig:fig4}(b-d), the exact calculation predicts a significant degradation of the polarization purity at increasing powers. At higher powers the quantum dot transition becomes saturated and the polarization purity is restored, as would be the case with an empty cavity. In Figs. ~\ref{fig:fig4}(b-d) we note the strong discrepancy between the exact calculation and the semiclassical approximation, which only predicts coherent outputs and, thus, pure polarization states. The semiclassical image should thus, by no means, be considered as a complete description of the optical response for a CQED device; it can be fruitfully used, however, in the absence of pure dephasing and for experiments in the low-excitation regime.

In summary, polarization tomography allows probing the optical response of a CQED device beyond the semiclassical approximation. We have shown that the superposition of $H$-polarized single photons with a $V$-polarized reflected field can be partially coherent, leading to a polarization rotation whose degree of purity can be decreased by dephasing processes. The tomography approach will provide a comprehensive measurement tool for polarization-encoded protocols: in particular, a single electron spin described in the Bloch sphere can be monitored by, or entangled with, a photon polarization qubit described in the Poincar\'e sphere. This would constitute the heart of future quantum networks, where CQED devices interact by exchanging polarization-encoded photonic qubits \cite{Hu2008}.



\acknowledgments

This work was partially supported by the ERC Starting Grant No. 277885 QD-CQED, the French Agence Nationale pour la Recherche (QDOM:  ANR-12-BS10-0010 and SPIQE: ANR-14-CE32-0012),  the Labex NanoSaclay, and the RENATECH network. 



%

\pagebreak
\widetext
\begin{center}
\textbf{\large Supplemental Materials: Tomography of optical polarization rotation induced by a quantum dot-cavity device}
\end{center}
\setcounter{equation}{0}
\setcounter{figure}{0}
\setcounter{table}{0}
\setcounter{page}{1}
\makeatletter
\renewcommand{\theequation}{S\arabic{equation}}
\renewcommand{\thefigure}{S\arabic{figure}}
\renewcommand{\bibnumfmt}[1]{[S#1]}
\renewcommand{\citenumfont}[1]{S#1}

\section{Section 1: Input-output formalism with polarized light}

Using the input-output formalism \cite{Walls2008}, we define $\hat{b}^{(in)}_V$ the operator describing the incident $V$-polarized field, $\hat{b}_V$ and $\hat{b}_H$ the operators describing the $H$- and $V$-polarized  output fields. We also define the dimensionless intracavity fields $\hat{a}_V$ and $\hat{a}_H$ for both polarizations. The continuity relations between these operators write: 
\begin{eqnarray} \label{eq_input_output}
\hat{b}_V & = & \hat{b}^{(in)}_V\:+\:\sqrt{\kappa_{top}} a_V \\
\hat{b}_H & = & \sqrt{\kappa_{top}} a_H,
\end{eqnarray}
with $\kappa_{top}$ the cavity intensity damping rate through the top mirror.\\

From these operators, one can define the output operators $\hat{b}_D$ and $\hat{b}_A$ in the diagonal/antidiagonal basis, $\hat{b}_L$ and $\hat{b}_R$ in the left-handed/right-handed circular basis:
\begin{eqnarray} \label{eq_operators_D_A_L_R}
\hat{b}_D & = & \left(\hat{b}_H\:+\:\hat{b}_V\right)\:/\:\sqrt{2} \\
\hat{b}_A & = & \left(\hat{b}_H\:-\:\hat{b}_V\right)\:/\:\sqrt{2} \\
\hat{b}_L & = & \left(\hat{b}_H\:-\:i\:\hat{b}_V\right)\:/\:\sqrt{2} \\
\hat{b}_R & = & \left(\hat{b}_H\:+\:i\:\hat{b}_V\right)\:/\:\sqrt{2} 
\end{eqnarray}

With these notations the intensity $I_i$, i.e. the number of output photons per unit time in polarization $i$ (with $i=H,V,D,A,L,R$), is given by  $I_i=\langle \hat{b}_i^{\dagger}\hat{b}_i\rangle$. This intensity is generally higher than $\langle \hat{b}_i^{\dagger}\rangle\,\langle\hat{b}_i\rangle$, which only describes the coherent part of the $i$-polarized output field.\\

\section{Section 2: Hamiltonian of the driven QD-cavity system}

We schematize in Fig.~S1(a) the three-level system of the neutral exciton confined in an anisotropic QD: the system eigenstates are the ground state $|g\rangle$ and two excited states spectrally separated by the QD fine structure splitting $\Delta_{FSS}$. These eigenstates $|e_X\rangle$ and $|e_Y\rangle$ can be optically addressed with $X$-polarized and $Y$-polarized light, respectively, $X$ and $Y$ being the eigenaxes of the QD anisotropy. In our experiment, however, the important axes are given by the cavity anisotropy, corresponding to vertical ($V$) and horizontal ($H$) polarizations: the $V$ and $H$ eigenaxes generally differ from the QD eigenaxes $X$ and $Y$ by an angle $\theta$, as illustrated in Fig.~S1(b) \cite{Giesz2016}.

\begin{figure}[ht!]\label{fig_S1}
\begin{center}
\includegraphics[width=0.7\linewidth,angle=0]{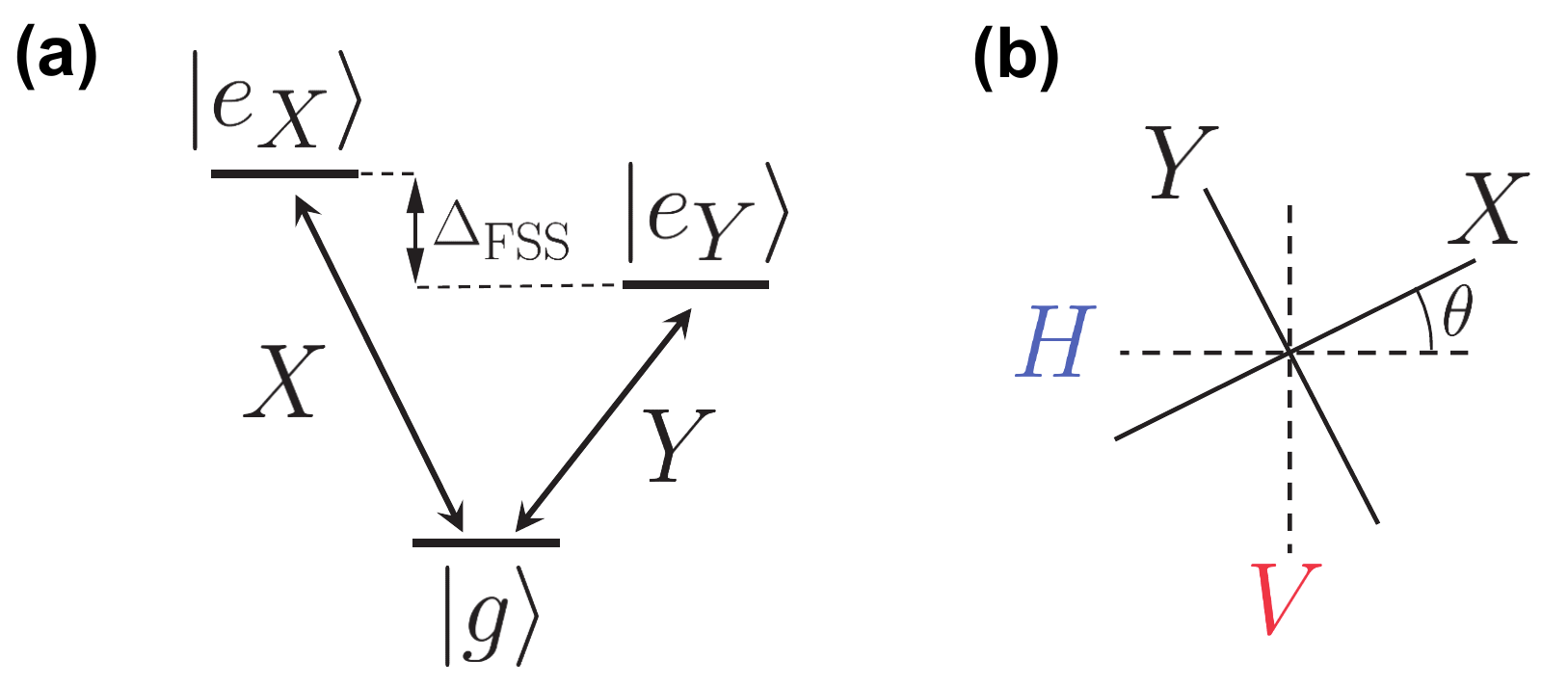}
\end{center}
\begin{flushleft}
FIG. S1.  \textbf{(a)} Quantum dot three level system. \textbf{(b)} Relative orientation of cavity and QD eigenaxes.
\end{flushleft}
\end{figure}

To describe the system we thus use the excitonic states $|e_H\rangle$ and $|e_V\rangle$ that can be optically addressed with $V$- and $H$-polarized light:
\begin{eqnarray}
|e_H\rangle & = & \: \: \  \cos(\theta) \: |e_X\rangle   \: + \: \sin(\theta) \: |e_Y\rangle \\
|e_V\rangle & = & - \sin(\theta) \: |e_X\rangle   \: + \: \cos(\theta) \: |e_Y\rangle
\end{eqnarray}
Denoting $\omega_{i}^{\textrm{\tiny{QD}}}$ the frequency of the QD state $|e_i\rangle$, with $i=X,Y,H,V$, we find that: 
\begin{eqnarray}
\omega_{H}^{\textrm{\tiny{QD}}} & = & \omega_{X}^{\textrm{\tiny{QD}}} \cos^2(\theta)+\omega_{Y}^{\textrm{\tiny{QD}}} \sin^2(\theta) \\
 \omega_{V}^{\textrm{\tiny{QD}}} & = & \omega_{X}^{\textrm{\tiny{QD}}} \sin^2(\theta)+\omega_{Y}^{\textrm{\tiny{QD}}} \cos^2(\theta)
\end{eqnarray} 
This allows writing the Hamiltonian $\hat{H}_{\textrm{QD}}$ of the QD system in the $(|e_H\rangle$, $|e_V\rangle)$ basis, in the frame rotating at the laser frequency $\omega$: 
\begin{eqnarray}
\hat{H}_{\textrm{QD}} & = & \left(\omega_{H}^{\textrm{\tiny{QD}}}-\omega\right)\hat{\sigma}_{H}^{+}\hat{\sigma}_{H}^{-}+\left(\omega_{V}^{\textrm{\tiny{QD}}}-\omega\right)\hat{\sigma}_{V}^{+}\hat{\sigma}_{V}^{-} \nonumber \\
 &  & \ \ \  \ \ + \: \Delta_{FSS}\cos\theta \sin\theta\left(\hat{\sigma}_{H}^{+}\hat{\sigma}_{V}^{-}+\hat{\sigma}_{V}^{+}\hat{\sigma}_{H}^{-}\right)
\end{eqnarray}
with $\hat{\sigma}_{i}^{-}$ the lowering operator $|g\rangle \langle e_i|$ and $\hat{\sigma}_{i}^{+}$ the corresponding raising operator.\\

We now turn to the Hamiltonian $\hat{H}_{\textrm{cav}}$ of the cavity system. Denoting $\omega_{i}^{c}$ the frequency of the $i$-polarized cavity mode, with $i=H,V$, we find:
\begin{equation}
\hat{H}_{\textrm{cav}} \: = \: \left(\omega_{H}^{c}-\omega\right)\hat{a}_{H}^{\dagger}\hat{a}_{H}+\left(\omega_{V}^{c}-\omega\right)\hat{a}_{V}^{\dagger}\hat{a}_{V}
\end{equation}
Denoting $b_{V}^{(in)}=\langle\hat{b}^{(in)}_V\rangle$, we also find the Hamiltonian $\hat{H}_{\textrm{pump}}$ describing the CW resonant excitation by the $V$-polarized input field:
\begin{equation}
\hat{H}_{\textrm{pump}} \: = \: -\: i \: \sqrt{\kappa_{top}} \: \left(b_{V}^{(in)}\:\hat{a}_{V}^{\dagger}\:-\:b_{V}^{(in)^*}\:\hat{a}_{V}\right)
\end{equation}
Finally, denoting $g$ the light-matter coupling strength, the light-matter interaction Hamiltonian $\hat{H}_{\textrm{LM}}$ writes:
\begin{equation}
\hat{H}_{\textrm{LM}} \:\:=\: -\: i\:g\:(\hat{a}_{H}^{\dagger}\hat{\sigma}_{H}^{-}\:+\:\hat{a}_{V}^{\dagger}\hat{\sigma}_{V}^{-}\:-\:\hat{\sigma}_{H}^{+}\hat{a}_{H}\:-\:\hat{\sigma}_{V}^{+}\hat{a}_{V})
\end{equation}

\section{Section 3: Dissipative processes - master equation}

We model the irreversible interaction between the cavity-QED system and its external environment  by a number of dissipative processes. Each process is described by a collapse operator $\hat{C}$, which participates to the evolution of the system density matrix $\rho$ via a Limbladian super-operator $\hat{L}$ \cite{Gardiner2004}:
\begin{equation}
\hat{L}\left[\rho\right]=\frac{1}{2}\left(2\hat{C}\rho\hat{C}^{\dagger}-\hat{C}^{\dagger}\hat{C}\rho-\rho\hat{C}^{\dagger}\hat{C}\right)
\end{equation}\\

The collapse operators associated to our dissipative processes are the following:
\begin{itemize}
\item \emph{Cavity damping}, associated to the cavity optical losses, is described by the collapse operators $
\hat{C}_i^{\textrm{cav}}=\sqrt{\kappa}\,\hat{a_i}$, with $i=H,V$ and $\kappa$ the total cavity damping rate.
\item \emph{Quantum dot spontaneous emission in the leaky modes}, associated to direct spontaneous emission of a photon into the external environment, is described by the collapse operators $\hat{C}_i^{\textrm{sp}}=\sqrt{\gamma_{sp}}\,\hat{\sigma}_{i}^{-}$, with $i=H,V$ and $\gamma_{sp}$ the QD spontaneous emission rate. 
\item \emph{Excitonic pure dephasing}, associated to the decoherence processes which preserve the population of the excited state $\left|e_H\right\rangle$ and $\left|e_V\right\rangle$, is described by a collapse operator $
\hat{C}_{\textrm{deph}}=\sqrt{2\gamma_{*}}\,\left(\hat{\sigma}_{H}^{+}\hat{\sigma}_{H}^{-}+\hat{\sigma}_{V}^{+}\hat{\sigma}_{V}^{-}\right)$, with $\gamma^{*}$ the pure dephasing rate.
\end{itemize}

These operators allow defining the corresponding Limbladian super-operators $\hat{L}_H^{\textrm{cav}}$, $\hat{L}_V^{\textrm{cav}}$, $\hat{L}_H^{\textrm{sp}}$, $\hat{L}_V^{\textrm{sp}}$ and $\hat{L}^{\textrm{deph}}$. Using the total system Hamiltonian $\hat{H}_{\textrm{tot}}=\hat{H}_{\textrm{QD}}+\hat{H}_{\textrm{cav}}+\hat{H}_{\textrm{pump}}+\hat{H}_{\textrm{LM}}$, we deduce the complete master equation describing the system evolution:
\begin{eqnarray}
\frac{d\rho}{dt} & = & -\: i\:[\hat{H}_{\textrm{tot}},\rho] \:+\:\hat{L}_H^{\textrm{cav}}\left[\rho\right]\:+\:\hat{L}_V^{\textrm{cav}}\left[\rho\right]\nonumber \\
& & \: +\:\hat{L}_H^{\textrm{sp}}\left[\rho\right]\:+\:\hat{L}_V^{\textrm{sp}}\left[\rho\right]\:+\:\hat{L}^{\textrm{deph}}\left[\rho\right]
\end{eqnarray}\\

Finally, when describing CW experiments this master equation is numerically solved to deduce the stationary state density matrix, $\rho_{ss}$, from which average values such as $\langle \hat{b}_i^{\dagger}\hat{b}_i\rangle$ can be deduced. The master equation can also be used to describe second-order correlations such as $\langle \hat{b}_H^{\dagger}(0) \hat{b}_H^{\dagger}(\tau) \hat{b}_H(\tau) \hat{b}_H \rangle(0)$, as was used for the numerical prediction of $g_H^{(2)}(\tau)$ in Fig. 2b (see main text).\\

\section{Section 4: Effect of residual uncoupled light}

Analyzing the reflected output from a cavity-QED device requires distinguishing between the light coupled into the cavity mode and the residual uncoupled light. We define $\eta_{in}$ the input-coupling efficiency, i.e. the overlap between the spatial profiles of the incoming beam and of the fundamental cavity mode. For any polarization $i$ the total input field intensity $I_i^{in}$ can then be divided in two contributions, namely $\eta_{in}I_i^{in}$ and $(1-\eta_{in})I_i^{in}$, respectively associated to coupled and uncoupled light. Similarly, the output intensity in a given polarization $i$ is the sum of $I_i^\mathrm{m}$, denoting the contribution from light coupled to the mode, and $I_i^\mathrm{\not{m}}$ for uncoupled light \cite{Loo2012}. These two contributions can be computed separately:

\begin{itemize}
\item \emph{Coupled light}. The intensity $I_i^\mathrm{m}$ in any polarization $i$ is described by $I_i^\mathrm{m}=\langle \hat{b}_i^{\dagger}\hat{b}_i\rangle$, directly deduced from the stationary density matrix $\rho_{ss}$, taking into account a coupled input intensity $\eta_{in} I_i^{in}$.
\item \emph{Uncoupled light}. The intensity $I_i^\mathrm{\not{m}}$ in any polarization $i$ is described by $I_i^\mathrm{\not{m}}=(1-\eta_{in})I_i^{in}$, i.e. the total amount of uncoupled light. Indeed, in our experimental configuration, the spatial profile of the incoming beam is slightly smaller than the micropillar top surface, allowing all the uncoupled light to be fully reflected by the optically flat surface. Furthermore, this reflection on the micropillar surface does not rotate the polarization of light, allowing $I_i^\mathrm{\not{m}}$ to be proportional to the input intensity $I_i^{in}$ for any polarization $i$.
\end{itemize}

In our experiment, the residual uncoupled light stays $V$-polarized and thus plays a role in the tomography measurement described in Fig. 3a; it also adds a constant contribution to the intensity $I_V$ described in Fig. 2c. The numerical fit presented in these figures takes this residual effect into account, with a best estimated value of $21\pm 5\%$ of uncoupled light. 

\end{document}